\documentclass[%
 reprint,
 amsmath,amssymb,
 aps,
prstper,
floatfix,
]{revtex4-2}

\usepackage{graphicx}
\usepackage{dcolumn}
\usepackage{bm}
\usepackage{hyperref}
\usepackage{xcolor}

\begin{document}

\preprint{APS/123-QED}

\title{
Thermalization in Quantum Fluids of Light: A Convection-Diffusion Equation 
}

\author{Vladislav Yu. Shishkov}
\email{vladislavmipt@gmail.com}
\affiliation{ Dukhov Research Institute of Automatics (VNIIA), 22 Sushchevskaya, Moscow 127055, Russia; }
\affiliation{ Moscow Institute of Physics and Technology, 9 Institutskiy pereulok, Dolgoprudny 141700, Moscow region, Russia; }

\author{Ivan V. Panyukov}
\affiliation{ Dukhov Research Institute of Automatics (VNIIA), 22 Sushchevskaya, Moscow 127055, Russia; }
\affiliation{ Moscow Institute of Physics and Technology, 9 Institutskiy pereulok, Dolgoprudny 141700, Moscow region, Russia; }

\author{Evgeny S. Andrianov}
\affiliation{ Dukhov Research Institute of Automatics (VNIIA), 22 Sushchevskaya, Moscow 127055, Russia; }
\affiliation{ Moscow Institute of Physics and Technology, 9 Institutskiy pereulok, Dolgoprudny 141700, Moscow region, Russia; }

\author{Anton~V.~Zasedatelev}
\email{anton.zasedatelev@univie.ac.at}
\affiliation{
Vienna Center for Quantum Science and Technology~(VCQ),
~Faculty~of~Physics,~University~of~Vienna, Boltzmanngasse~5, 1090~Vienna, Austria
}

\date{\today}

\begin{abstract}

We develop a microscopic theory for the dynamics of quantum fluids of light, deriving an effective kinetic equation in momentum space that takes the form of the convection-diffusion equation.
In the particular case of two-dimensional systems with parabolic dispersion, it reduces to the Bateman--Burgers equation. 
The hydrodynamic analogy unifies nonlinear wave phenomena, such as shock wave formation and turbulence, with non-equilibrium Bose--Einstein condensation of photons and polaritons in optical cavities. 
We introduce the Reynolds number $(\textit{Re})$ and demonstrate that the condensation threshold corresponds exactly to a critical Reynolds number of unity $(\textit{Re}=1)$, beyond which $(\textit{Re} > 1)$ a shock-like front emerges in the momentum space, characterized by the Bose--Einstein distribution for the particle density in states with high momentum.


\end{abstract}

\maketitle

\section{Introduction}

Quantum fluids of light exhibit remarkable collective behaviors driven by nonlinear interactions between photons or polaritons, giving rise to a diverse array of hydrodynamical phenomena~\cite{carusotto2013quantum}.
Thermalization of the particles is a crucial process in the formation of light-matter Bose--Einstein condensates (BECs)~\cite{kirton2013nonequilibrium, kirton2015thermalization, schmitt2015thermalization}. In optical cavities, this process depends on the absorption and emission of photons by dye molecules within the cavity. In this case, the rate of energy exchange between the cavity photons and the molecules, as well as their vibrational properties, significantly impacts the efficiency of thermalization. 
In this study, we develop an effective equation that describes the kinetics of particles over momentum space, resembling the Bateman--Burgers equation~\cite{whitham2011linear, orlandi2000fluid} known from fluid dynamics.
We introduce the concept of the Reynolds number $(\textit{Re})$ for the two-dimensional Bose gas and show that the critical value of this number $(\textit{Re}=1)$ not only marks the onset of significant nonlinear dynamics, but also coincides with the threshold for Bose--Einstein condensation. 

\section{Kinetics of the Bose gas}
The dynamics of the average number of particles, $n_{\bf k}(t)$, with momentum $\bf k$ is governed by three processes: thermalization, dissipation with the rate $\gamma_{\bf k}$, and incoherent pumping with the rate $\varkappa_{\bf k}(t)$ which may depend on time.
Generally, the kinetics of the Bose gas can be described by the Maxwell--Boltzmann equation~\cite{deng2010exciton}.
\begin{multline} \label{Maxwell-Boltzmann}
\frac{\partial n_{\bf k}(t)}{\partial t} =
-\gamma_{\bf k} n_{\bf k}(t) 
+
\sum_{\bf k'}
\left[
\gamma_{\rm therm}^{\bf k k'}\left( n_{\bf k}(t) + 1\right)n_{\bf k'}(t) -
\right.
\\
\left.
\gamma_{\rm therm}^{\bf k' k}\left( n_{\bf k'}(t) + 1\right)n_{\bf k}(t)
\right] + \varkappa_{\bf k}(t),
\end{multline} 
An important property of the particles affecting their kinetics is their dispersion, $\omega_{\bf k}$, that establishes the connection between the momentum of the particle, $\hbar {\bf k}$, and its energy, $\hbar \omega_{\bf k}$.
We assume that (1) the system is isotropic, that is, the energy of the particles and the dissipation rate depend only on the absolute value of the momentum, $k$, i.e. $\hbar \omega_{\bf k} = \hbar \omega_{k}$ and $\gamma_{\bf k} = \gamma_{k}$; (2) $\omega_{k}$ increases monotonically with $k$.
These assumptions are relevant for most experiments on Bose--Einstein condensation.

Thermalization part of the equation causes the redistribution of the particles in the momentum space and preserves the total number of particles $\sum_{\bf k} n_{\bf k}(t)$.
Indeed, from~(\ref{Maxwell-Boltzmann}) it follows
\begin{equation} \label{conservation of the particles}
\frac{d}{dt}\sum_{\bf k} n_{\bf k}(t) = - \sum_{\bf k} \gamma_{\bf k} n_{\bf k}(t) + \sum_{\bf k} \varkappa_{\bf k}(t).
\end{equation}
Regarding the thermalization process, we assume that (1) the thermalization rate is zero between the states separated by the energies over $\hbar \omega_M$, i.e. $\gamma_{\rm therm}^{{\bf k}_1 {\bf k}_2} = 0$ for $|\omega_{k_1}-\omega_{k_2}|>\omega_M$; (2) the thermalization rate depends only on the difference between the energies of the states such that
\begin{multline} \label{assumption 1}
\gamma_{\rm therm}^{{\bf k}_1 {\bf k}_2} = \gamma_{\rm therm}^{{k}_1 {k}_2} = \Gamma \left( 1 + n_{\rm th} (\omega_{k_2}-\omega_{k_1}) \right)
\\ \text{for}\;\; |\omega_{k_1}-\omega_{k_2}|<\omega_M  \; \text{and} \; \omega_{k_1}<\omega_{k_2},
\end{multline}
and
\begin{multline} \label{assumption 2}
\gamma_{\rm therm}^{{\bf k}_1 {\bf k}_2} = \gamma_{\rm therm}^{{k}_1 {k}_2} = \Gamma n_{\rm th} (\omega_{k_1}-\omega_{k_2})
\\ \text{for}\;\; |\omega_{k_1}-\omega_{k_2}|<\omega_M  \; \text{and} \; \omega_{k_2}<\omega_{k_1},
\end{multline}
where $\Gamma$ is a constant rate, $n_{\rm th} (\Delta\omega) = (e^{\hbar \Delta\omega/k_BT} - 1)^{-1}$, $T$ is the temperature of the environment, and $k_B$ is the Boltzmann constant.
These assumptions are consistent with the Kubo--Martin--Schwinger relation~\cite{kubo1957statistical}, $\gamma_{\rm therm}^{{\bf k}_1 {\bf k}_2} = \gamma_{\rm therm}^{{\bf k}_2 {\bf k}_1} e^{-(\hbar\omega_{{\bf k}_1} - \hbar\omega_{{\bf k}_2})/k_BT}$.

\begin{figure}
\includegraphics[width=0.9\linewidth]{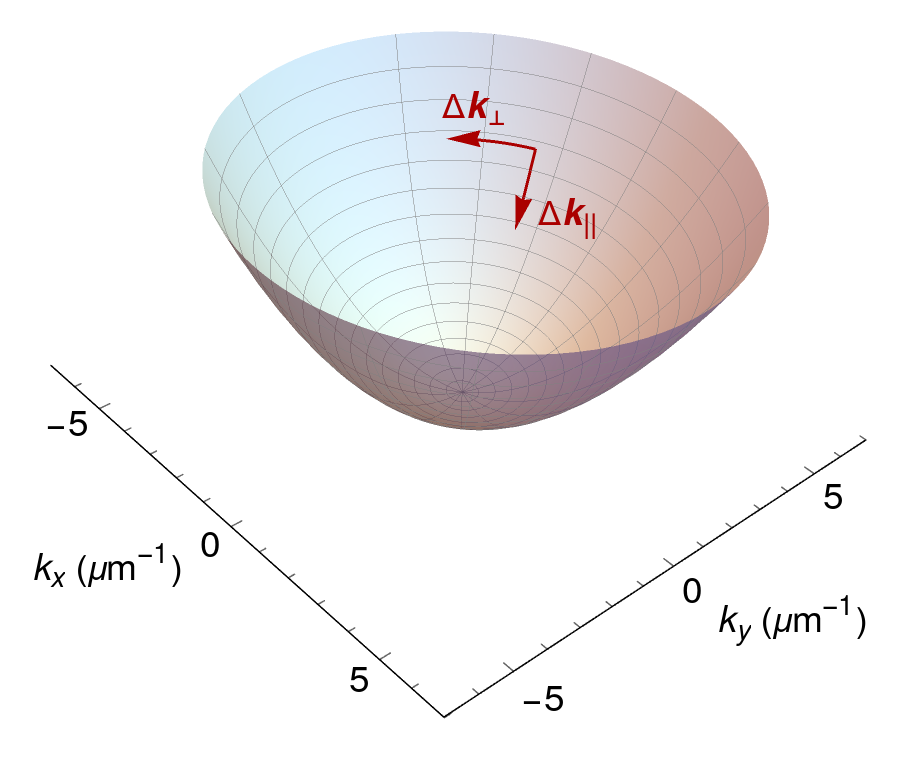}
\caption{
Energy-momentum relation of the particles.
For the illustrative purposes we show a two-dimensional system with the particles having a parabolic dispersion.
For the illustrative purposes, we also specify the momentum range from $-7$~$\mu$m$^{-1}$ to $7$~$\mu$m$^{-1}$ which is relevant for the recent experiments~\cite{sannikov2024room, plumhof2014room, zasedatelev2019room, zasedatelev2021single}.
Red arrows on the dispersion surface show perpendicular $\Delta {\bf k}_{\perp}$ and parallel $\Delta {\bf k}_{||}$ components of the wave-vector, exchanged within the process of particle thermalization.
}
    \label{fig:dispersion}
\end{figure}

The momentum $\bf k$ in Eq.~(\ref{Maxwell-Boltzmann}) are discrete, and their manifold is determined by the size of the system. We make a transition from the discrete manifold to the continuous distribution of the momentum introducing the linear density of the particles
\begin{equation} \label{linear density}
f(k,t) = \sum_{\bf K} n_{\bf K}(t) \delta(|{\bf K}| - k),    
\end{equation}
where the sum is taken over all the states with the same absolute momentum $\textbf{k}$.
The introduction of the linear density~(\ref{linear density}) allows us not only to pass from discrete momenta to a continuous momentum $k$ in Eq.~(\ref{Maxwell-Boltzmann}), but also exclude thermalization dynamics along the states with perpendicular momenta, as shown in Fig.~\ref{fig:dispersion}.

One can see that the thermalization rate given by~(\ref{assumption 1})-(\ref{assumption 2}) does not depend on $\Delta {\bf k}_\perp$ (see~Fig.~\ref{fig:dispersion}).
Thus, using the definition of the linear density~(\ref{linear density}), we can transform the discrete Maxwell--Boltzmann equation~(\ref{Maxwell-Boltzmann}) into the continuous form
\begin{multline} \label{integral eq}
\frac{\partial f(k,t)}{\partial t} =
-\gamma_k f(k,t)
+
F(k,t)
\\
+ 
\int\limits_0^{+\infty}
(\gamma^{kk'}_{\rm therm}-\gamma^{k'k}_{\rm therm}) f(k,t)f(k',t)dk' 
\\ 
+
\int\limits_0^{+\infty} 
\left[\gamma^{kk'}_{\rm therm}g(k)f(k',t)-\gamma^{k'k}_{\rm therm}g(k')f(k,t)\right]
dk'
.
\end{multline}
where $F(k,t) = \sum_{\bf K} \varkappa_{\bf K}(t) \delta(|{\bf K}| - k)$ is an incoherent pumping and $g(k)=\sum_{\bf K} \delta(|{\bf K}| - k)$ is the density of states determined by the dimensionality of the system.
For a one-dimensional system $g(k)=L/2\pi$, where $L$ is the length of the system; for a two-dimensional system $g(k)=Sk/2\pi$, where $S$ is the area of the system; for a two-dimensional system $g(k)=Vk^2/2\pi^2$, where $V$ is the volume of the system.

So far, Eq.~(\ref{integral eq}) is equivalent to Eq.~(\ref{Maxwell-Boltzmann}) which is the Maxwell--Boltzmann equation.
To transform the integral equation~(\ref{integral eq}) into the equation of a differential type, we  integrate over the momentum space preserving only the terms up to $\omega_M^2$ (see Appendix~\ref{appendix: derivation} for details).
As a result, we obtain the convection-diffusion equation describing the kinetics of Bose gas 
\begin{equation}  \label{eq for linear density}
\frac{\partial f(k,t)}{\partial t} + \frac{\partial J(k,t)}{\partial k} 
=
-\gamma_k f(k,t) + F(k,t)
\end{equation}
with the flux in momentum space defined by
\begin{multline} \label{flow of the particles}
J(k,t) 
= 
-
\frac{\Gamma\omega_M^2}{2} 
\frac{g(k)f(k,t)}{v_{\rm gr}^2(k)}
\\
-
\frac{k_BT}{\hbar}
\frac{\Gamma\omega_M^2}{2}
\frac{g^2(k)}{v_{\rm gr}^3(k)} 
\frac{\partial }{\partial k} 
\left(
\frac{f(k,t)}{g(k)}
\right)
-
\frac{\Gamma\omega_M^2}{2}
 \frac{f^2(k,t)}{v_{\rm gr}^2(k)}.
\end{multline}
The left-hand side of Eq.~(\ref{eq for linear density}) corresponds to the conservative dynamics with the particle redistribution over momentum space caused by thermalization, while the right-hand side of this equation is the non-conservative part, which describes the loss of particles, $-\gamma_k f(k,t)$, and the supply of particles to the system from some external source, $F(k,t)$.
Since $k = |{\bf k}|$, the particles cannot pass through $k=0$ implying the boundary condition for Eq.~(\ref{eq for linear density}) 
\begin{equation} \label{boundary condition J}
{\lim_{k\to+0}}J(k,t) = 0.
\end{equation}

The convection–diffusion equation in momentum space~(\ref{eq for linear density}) with boundary condition~(\ref{boundary condition J}) and the expression for the flow of the particles in momentum space~(\ref{flow of the particles}) are the main result of the paper.
In what follows, we study the properties of the obtained equation with the main focus on two-dimensional systems with a parabolic dispersion.

\section{Ganeral properties of the convection–diffusion equation (\ref{eq for linear density})}

Similar to the Maxwell–Boltzmann equation~(\ref{Maxwell-Boltzmann}), the thermalization term, $ {\partial J(k,t)/\partial k} $, in Eq.~(\ref{eq for linear density}) contains both linear component (the first two terms on the right-hand side of Eq.~(\ref{flow of the particles})) and a non-linear component (the last term on the right-hand side of Eq.~(\ref{flow of the particles})).
However, unlike the Maxwell–Boltzmann equation~(\ref{Maxwell-Boltzmann}), the equation governing the linear density~(\ref{eq for linear density}) is local with respect to $k$, which facilitates both analytical and numerical analysis.
Notably, the thermalization property of the Maxwell–Boltzmann equation~(\ref{Maxwell-Boltzmann}), which ensures conservation of the total number of particles, holds for Eq.~(\ref{eq for linear density}).
Indeed, from Eq.~(\ref{linear density}) and (\ref{eq for linear density}), it follows $\int_{0}^{+\infty}f(k, t)dk = \sum_{\bf k}n_{\bf k}(t)$ and $(\partial/\partial t) \int_0^{+\infty}f(k, t)dk = -\gamma_k\int_0^{+\infty}f(k, t)dk + \int_0^{+\infty}F(k,t)dk$ which is equivalent to Eq.~(\ref{conservation of the particles}).

In the conservative scenario, $\gamma_k=0$ and $F(k,t)=0$ in Eq.~(\ref{eq for linear density}), the stationary linear density of the particles, $f_{\rm St}(k)$, follows the equation $J = 0$, thus, from~Eq.~(\ref{flow of the particles}) we have
\begin{equation} \label{eq for linear density stationary}
\frac{g(k)f_{\rm St}(k)}{v_{\rm gr}^2(k)} 
+ 
\frac{k_BT}{\hbar}
\frac{g^2(k)}{v_{\rm gr}^3(k)} 
\frac{\partial }{\partial k} 
\left(
\frac{f_{\rm St}(k)}{g(k)}
\right)
+
\frac{f_{\rm St}^2(k)}{v_{\rm gr}^2(k)}
=0
\end{equation}
It is easy to see, that the linear density
\begin{equation}
f_{\rm St}(k) = \frac{g(k)}{e^{(\hbar (\omega_k-\omega_{\bf 0}) - \mu)/k_BT} - 1},
\end{equation}
corresponding to the Bose--Einstein distribution, $n_{\bf k} = (e^{(\hbar (\omega_{\bf k}-\omega_{\bf 0}) - \mu)/k_BT} - 1)^{-1}$ with the chemical potential $\mu$, is the solution of Eq.~(\ref{eq for linear density stationary}).
Thus, in the conservative scenario, Eq.~(\ref{eq for linear density}) reproduces Bose--Einstein distribution as the steady-state solution.

\section{2D Bose gas with parabolic dispersion}
For a 2D system with the area $S$, as it was mentioned above, $g(k) = Sk/2\pi$.
The parabolic dispersion implies $\omega_{k} = \omega_{\bf 0} + \alpha k^2$, consequently, $v_{\rm gr}(k) = 2 \alpha k$.
As a result, from Eq.~(\ref{eq for linear density}) we obtain
\begin{multline} \label{eq for linear density 2D}
\frac{\partial f(k,t)}{\partial t}
=
-\gamma_k f(k,t)
+ \frac{v}{2} \frac{\partial}{\partial k} \left( \frac{f(k,t)}{k} \right)
+
\\
\frac{D}{4}\frac{\partial }{\partial k} 
\left( 
\frac{1}{k} 
\frac{\partial }{\partial k}
\left( 
\frac{f(k,t)}{k} 
\right)
\right)
+\frac{b}{8\pi} \frac{\partial }{\partial k} \left( \frac{f^2(k,t)}{k^2} \right)
+
F(k,t),
\end{multline}
where we introduced the coefficients
\begin{equation} \label{D}
D = 
\frac{k_BT}{\hbar}
\frac{S}{8\pi\alpha^3}
\Gamma\omega_M^2
,
\end{equation}
\begin{equation} \label{v}
v = 
\frac{S}{8\pi\alpha^2} \Gamma \omega_M^2
,
\end{equation}
\begin{equation} \label{b}
b = 
\frac{\pi}{\alpha^2}
\Gamma\omega_M^2
.
\end{equation}
A very important property of these coefficients is that the ratios between them depend only on the size of the system~$S$, dispersion coefficient~$\alpha$, and the temperature~$T$. 


\begin{figure*}
\includegraphics[width=1\linewidth]{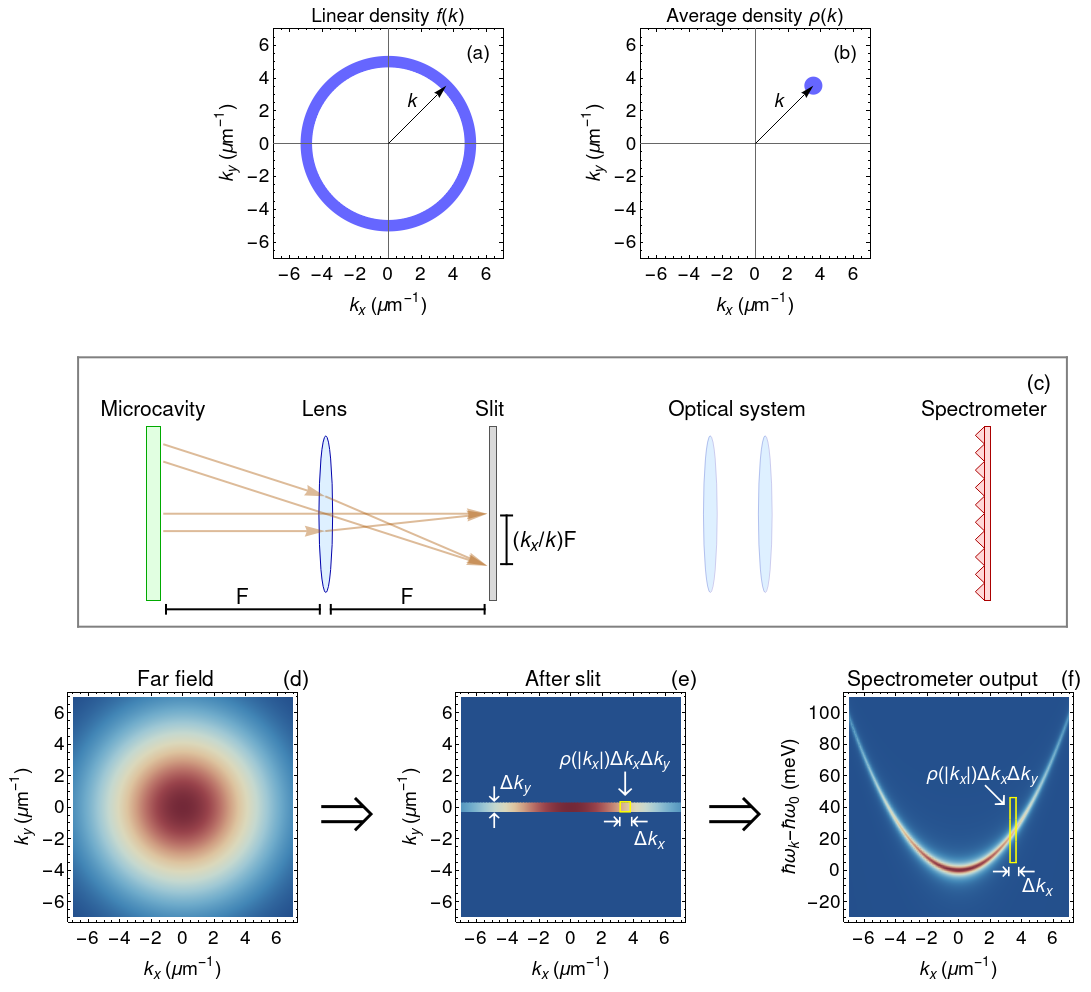}
\caption{
Schematic of (a) the linear density~(\ref{linear density}) and (b) the average density~(\ref{average density}) for a two-dimensional system.
Typical experimental setup to access average particle density (c).  The light from the cavity is collected in the confocal configuration resulting in far field (Fourier) image - momentum space (d).
The image is then projected to a slit of the spectrometer, $k_y=0$ as it is shown in (e).
Each pixel column at the CCD camera corresponds to frequency and each pixel row corresponds to the in-plane momentum of the emitted light.
The illustration of the resultant data is shown in (f).
We set the parameters as follows $\hbar\alpha_{\rm cav} = 2~{\rm (meV\cdot\mu m^2)}$, $\hbar\gamma_k = 5~{\rm (meV)}$ for all the cavity modes which is relevant for the recent experiments~\cite{sannikov2024room, zasedatelev2019room, zasedatelev2021single}.
}
    \label{fig: experimental setup}
\end{figure*}

\subsection{Bateman--Burgers equation}

While the linear density of particles~(\ref{linear density}) is quite convenient for theoretical analysis, it appears to be not very practical quantity for experimental observation of two-dimensional systems. Instead, the average density, $\rho(k, t)$, defined by
\begin{equation} \label{average density}
\rho(k, t) = \frac{f(k,t)}{2\pi k}
\end{equation} 
can be easily accessed in standard dispersion imaging measurements, schematically shown in Fig.~\ref{fig: experimental setup}.
Specifically, Fig.~\ref{fig: experimental setup}c shows the typical experimental setup to measure particle density in energy and momentum space. Here, the light from the cavity is collected in the confocal configuration, and then projected onto the slit of the spectrometer~\cite{jiang2022exciton, kasprzak2006bose, nitsche2016spatial, vakevainen2020sub, tang2021room, zasedatelev2019room, zasedatelev2021single}.
The lens produces a far-field (Fourier) image in the focal plane (Fig.~\ref{fig: experimental setup}d). The image is then projected on the entrance slit of the spectrometer such that only a narrow region around $k_y=0$ is coupled in (Fig.~\ref{fig: experimental setup}d).
Symmetric projection of the momentum space onto the slit ensures that the total number of particles within the region of interest (yellow square in Fig.~\ref{fig: experimental setup}e) is proportional to the average density~(\ref{average density}).
By integrating the emission over the frequencies within interval $\Delta k_x$ around $k_x$ (yellow rectangular in Fig.~\ref{fig: experimental setup}f) one can immediately obtain the average particle density $\rho(|k_x|,t)$ \ref{average density}.
Now, using the resultant equation~(\ref{eq for linear density 2D}) for the linear density, $f(k,t)$, we can get an equation for the evolution of the average particles density 
\begin{equation} \label{eq for average density}
\frac{\partial \rho}{\partial t}
=
-\gamma_k \rho
+v \frac{\partial \rho}{\partial (k^2)} 
+D\frac{\partial^2 \rho}{\partial (k^2)^2}
+b \rho\frac{\partial \rho}{\partial (k^2)} 
+ P
\end{equation}
where $\rho=\rho(k,t)$ and $P(k,t) = F(k,t)/2 \pi k$ is an incoherent pumping term.
The general boundary condition Eq.~(\ref{boundary condition J}) corresponds to the specific boundary condition for the average density of the particles in two-dimensional system with parabolic potential
\begin{equation} \label{boundary condition}
v \rho(0, t) 
+ 
D \frac{\partial\rho(0, t)}{\partial (k^2)}
+
\frac{b}{2} \rho^2(0, t)
=
0.
\end{equation}
Eq.~(\ref{eq for average density}) is very similar to the well-known Bateman--Burgers equation, which appears in fluid mechanics~\cite{orlandi2000fluid} describing turbulent flows, as well as the formation of shock waves~\cite{burgers1948mathematical}. 
We identify three major deviations from the standard Bateman--Burgers dynamics, namely: (i) -- the dissipation term $\gamma_k\rho$ corresponding to the loss of particles with the rate $\gamma_k$; (ii) -- the drift term $v \partial \rho/\partial(k^2)$ with the constant velocity $v$ towards the ground state $k=0$; and (iii) -- the semi-boundness of the momentum space ($k>0$) which leads to the boundary condition~(\ref{boundary condition}).

Following, the hydrodynamic analogy we introduce the Reynolds number, $Re$, by the ratio of the inertia forces (non-linear term $(b/2) {\partial \rho^2}/{\partial (k^2)}$) to viscous forces (diffusion term $D{\partial^2 \rho}/{\partial (k^2)^2}$)~\cite{anderson2011ebook, rathakrishnan2013theoretical} 
\begin{equation} \label{Reynolds}
Re =
\left[
\frac{b\rho^2}{4k\Delta k}
\right]
\cdot
\left[
\frac{D\rho}{(2k\Delta k)^2}
\right]^{-1}
=
\frac{N}{G_{2D}}
\end{equation}
where $\Delta k$ is the characteristic scale of the particles distribution over absolute momentum, $G_{2D} = {S k_B T}/{4\pi \hbar \alpha }$ is the number of states within the energy region $(\hbar\omega_{\bf 0},~\hbar\omega_{\bf 0} + k_BT)$~\cite{shishkov2022analytical}, and $N=\sum_{\bf k}n_{\bf k}\approx2\pi k \Delta k\rho(k)$ is the total number of particles.
Large Reynolds number, $Re \gg 1$, corresponds to strong non-linear dynamics, whereas small Reynolds number, $Re \ll 1$, describes the linear regime of the density evolution in the momentum space, comprising the diffusion and drift towards $k=0$. When the Reynolds number approaches unity, $Re = 1$, the system undergoes a transition to non-equilibrium Bose--Einstein condensation~\cite{shishkov2022analytical}, establishing the critical threshold within the hydrodynamical formalism.


\subsection{Solution of Bateman--Burgers equation}
Given that a high-momentum state is initially occupied -- created, for example, by an external seed beam injecting particles into the system at $k_{\rm seed}$ -- we can directly determine the time $t_0$ required for the population to reach the ground state using Eq.~(\ref{eq for average density}). In the regime below condensation threshold, where  $Re < 1$, we have mainly two terms: the drift and the diffusion, which leads to
\begin{equation} \label{migration time}
t_0 = 
\frac{k_{\rm seed}^4}{2 D}
\left(
\frac{1}{2} + \frac{\hbar \alpha k^2_{\rm seed}}{2k_BT}
+
\sqrt{\frac{1}{4} + \frac{\hbar \alpha k^2_{\rm seed}}{2k_BT}}
\right)^{-1}
\end{equation}
Therefore, in order for particles to reach the ground state, it is necessary that $t_0 < \gamma_{k=0}^{-1}$.
When $\hbar \alpha k^2_{\rm seed} \lesssim k_BT$, Eq.~(\ref{migration time}) results in $t_0 \propto T^{-1}$, which is in agreement with experimental observations of thermalization in exciton-polariton systems at room temperature~\cite{plumhof2014room}.

Upon reaching the ground state, the thermalization dynamics undergoes a significant change.
To tackle this problem, we analyze the conservative limit ($\gamma_k=0$) of Eq.~(\ref{eq for average density}), which reduces to the Bateman--Burgers equation with drift term in a semi-bounded space.
We obtain the stationary solution for the average particle density (see Appendix~\ref{appendix: stationary})
\begin{equation} \label{stationary solution}
\rho_{\rm St}(k) 
=
\frac{S}{4\pi^2}
\frac{1}{e^{(\hbar\alpha k^2-\mu)/k_BT}-1},
\end{equation}
where $\mu$ is the chemical potential of the particles.
Remarkably, the stationary solution of the Bateman--Burgers dynamics of a quantum fluid of light in 2D leads to the Bose--Einstein distribution with Eq.~(\ref{eq for average density}) representing the dynamics of the system in momentum space.

\begin{figure}[h!]
\includegraphics[width=1\linewidth]{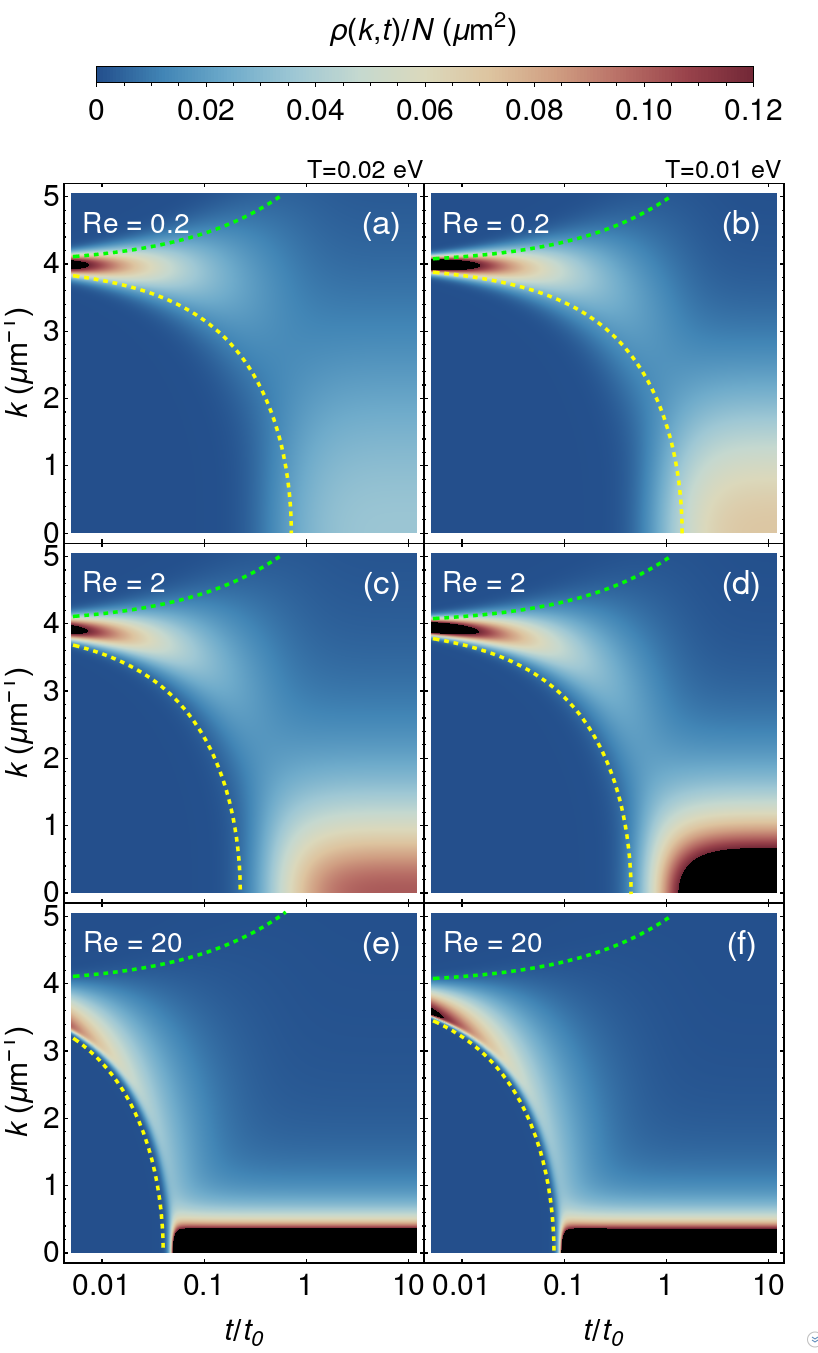}
\caption{
Conservative dynamics for the average particle density given by Eq.~(\ref{eq for average density}) at the different Reynolds numbers~(\ref{Reynolds}).
The time is normalized to the thermalization time, $t_0$, defined by Eq.~(\ref{migration time}), whereas the average particle density is normalized to the total density, $N$.
The Reynolds numbers are marked on the figures, the corresponding total number of particles is the following: (a)~$N = 159$, (b)~$N = 80$, (c)~$N = 1592$, (d)~$N = 798$, (e)~$N = 15916$, (f)~$N = 7958$.
Black regions mark the average densities of the photons surpassing $0.12 N~{\rm \mu m^{2}}$.
The yellow dashed lines mark the leading front, $k_{\rm front}(t)$, given by~Eq.~(\ref{k front}) and the green dashed lines back edge, $k_{\rm back}(t)$, given by~Eq.~(\ref{k back}). The parameters of the system are $S=1000~{\rm \mu m^2}$, $\hbar\alpha = 2~{\rm meV\cdot\mu m^2}$, $k_{\rm seed}=4~{\rm \mu m^{-1}}$, (a, c, e) $k_BT = 20~{\rm meV}$, (b, d, f) $k_BT = 10~{\rm meV}$, in agreement with experimental settings for polariton condensates~\cite{plumhof2014room, zasedatelev2021single}}
    \label{fig: conservative dynamics}
\end{figure}

Studying the dynamics of the average particle density $\rho(k,t)$ described by Eq.~(\ref{eq for average density}) requires a numerical approach. However, the conservative limit of Eq.~(\ref{eq for average density}) (with $\gamma_k = 0$ and $P = 0$) can still be solved analytically. For this case, we derive an exact solution that fulfills the boundary condition~(\ref{boundary condition}) (see Appendix~\ref{appendix: exact})
\begin{multline} \label{exact solution}
\rho(k,t) 
= 
2\frac{D}{b}
\frac{\partial}{\partial (k^2)}
\ln
\Bigg[
e^{-v k^2/2D}
\Bigg.
\\
\Bigg.
\int\limits_0^{+\infty}
G(k, t, k_1)
e^{(v k_1^2/2D)}
e^{\int^{k_1^2} \rho(k_2,0) d(k_2^2)}
d (k_1^2)
\Bigg]
\end{multline}
where
\begin{multline} \label{propagator}
G(k, t, k_1) = \\
\frac{e^{-[(k^2+k_1^2)^2+v^2t^2]/4Dt}}{\sqrt{4\pi D t}}  -
\frac{e^{-[(k^2-k_1^2)^2+v^2t^2]/4Dt}}{\sqrt{4\pi D t}} -
\delta(k_1^2)
\\
\left[
\frac{e^{v^2k^2/4D}}{2}
\Phi\left(\frac{k^2+vt}{\sqrt{4Dt}}\right)
+
\frac{e^{-v^2k^2/4D}}{2}
\Phi\left(\frac{k^2-vt}{\sqrt{4Dt}}\right)
\right]
\end{multline}
and $\Phi(\xi)$ is the complementary error function.

We use this result~(\ref{exact solution}) to capture the main features of the thermalization dynamics of quantum fluids of light. Our setup is based on the preoccupied high-momentum states, which can be seeded optically ${\bf k}_{\rm seed}$~\cite{yoon2022enhanced, sannikov2024room}.
Following this setup, we consider an initial condition 
\begin{equation} \label{particular initial condition}
\rho(k,0) = Re \cdot \frac{2D}{b} \delta(k^2-k_{\rm seed}^2)
\end{equation}
where $Re$ is the Reynolds number.
This initial condition corresponds to $N = 2\pi Re D/b$, consistent with $N = G_{2D}\cdot Re$ obtained in~Eq.~(\ref{Reynolds}). 
The exact solution for the average particle density in time and in momentum space given initial conditions~(\ref{particular initial condition}) can be found as follows
\begin{equation} \label{exact solution special case}
\rho(k,t) =
\frac{2v}{b}
\frac{
( e^{-\xi_+^2}/\sqrt{\pi v^2t/D} )
(1+e^{k^2k_{\rm seed}^2/Dt})
+ 
{\rm \Phi}(\xi_{+})
}{
e^{vk^2/D}
\left[2 (1-e^{-Re})^{-1} - {\rm \Phi}(\xi_{-})\right] - {\rm \Phi}(\xi_{+})
}
\end{equation}
where $\xi_{\pm} = (k^2\pm (k^2_{\rm seed} - vt))/\sqrt{4 D t}$.
The steady-state solution ($t\to +\infty$ ) for the average particle density reproduces Bose--Einstein distribution~(\ref{stationary solution}) with chemical potential $\mu = k_BT \ln(1-e^{-Re})$.

Figure~\ref{fig: conservative dynamics} showcases the thermalizaton dynamics for quantum fluids of light at different Reynolds numbers, $Re$.
In the conservative scenario, the population thermalizes to the ground state $k=0$ regardless of the Reynolds number. 
For small and intermediate Reynolds numbers, $Re \lesssim 1$, (Fig.~\ref{fig: conservative dynamics}a-b), thermalization is linear (drift and diffusion)
\begin{equation}
\rho(k,t) \approx
\frac{N}{2\pi}
\frac{e^{-(k^2-k^2_{\rm seed} + vt)^2/4Dt}}{\sqrt{\pi D t}}
\end{equation}
and the population reaches the ground state in time $t_0$, defined by Eq.~(\ref{migration time}).
For $Re > 1$ thermalization dynamics becomes nonlinear, speeding up the process by $t_0/Re$ approximately.
For $Re \gg 1$ and $t \lesssim t_0/Re$ we observe the abrupt change in the average particle density at the leading front (Fig.~\ref{fig: conservative dynamics}) resembling a shock wave in fluid mechanics~\cite{orlandi2000fluid}.
In this case, the momentum of the leading edge of the shock wave, $k_{\rm front}(t)$, and the tailing edge of the shock wave, $k_{\rm back}(t)$, are 
\begin{equation} \label{k front}
k_{\rm front}^2(t) = k_{\rm seed}^2 - \sqrt{\pi Re D t} - \sqrt{2 D t},
\end{equation}
\begin{equation} \label{k back}
k_{\rm back}^2(t) = k_{\rm seed}^2 + \sqrt{2 D t},
\end{equation}
Eq.~(\ref{k back}) stands for the diffusion of the tailing edge of the shock wave, whereas Eq~(\ref{k front}) reveals non-trivial dynamics for the leading edge.
First, thermalization in the non-linear regime accelerates by a factor of $Re$.
Second, the leading edge of the shock wave also diffuses, shifting the shock front by $\sqrt{2 D t}$ toward $k = 0$.
Both the leading front $k_{\rm front}$ and the back edge $k_{\rm back}$ are shown in Fig.~\ref{fig: conservative dynamics}e,f.
They not only closely follow the leading and tailing edges of the shock wave ($Re \gg 1$) but also describe the kinetics of the Bose gas for lower Reynolds numbers (Fig.~\ref{fig: conservative dynamics}a-d).

The density inside the shock wave ($Re \gg 1$) has the form $\rho(k,t) \approx B(t)\cdot(k_{\rm back}(t)-k)$ for $k_{\rm front}(t) \leq k \leq k_{\rm back}(t)$ and $\rho = 0$ otherwise with a coefficient $B(t)$ analogous to Ref.~\cite{orlandi2000fluid}.
We define $B(t)$ such that the total particle number, $N$, is conserved in time and obtain the approximate expression for the evolution of the average particle density for $k_{\rm front}(t) \leq k \leq k_{\rm back}(t)$
\begin{equation}
\rho(k,t) \approx 
\frac{3N (k_{\rm back}(t)-k)}{\pi(k_{\rm back}(t)-k_{\rm front}(t))^2(2k_{\rm front}(t)+k_{\rm back}(t))}
\end{equation}
which is valid as long as $k_{\rm front}(t) > 0$.

\begin{figure}
\includegraphics[width=0.85\linewidth]{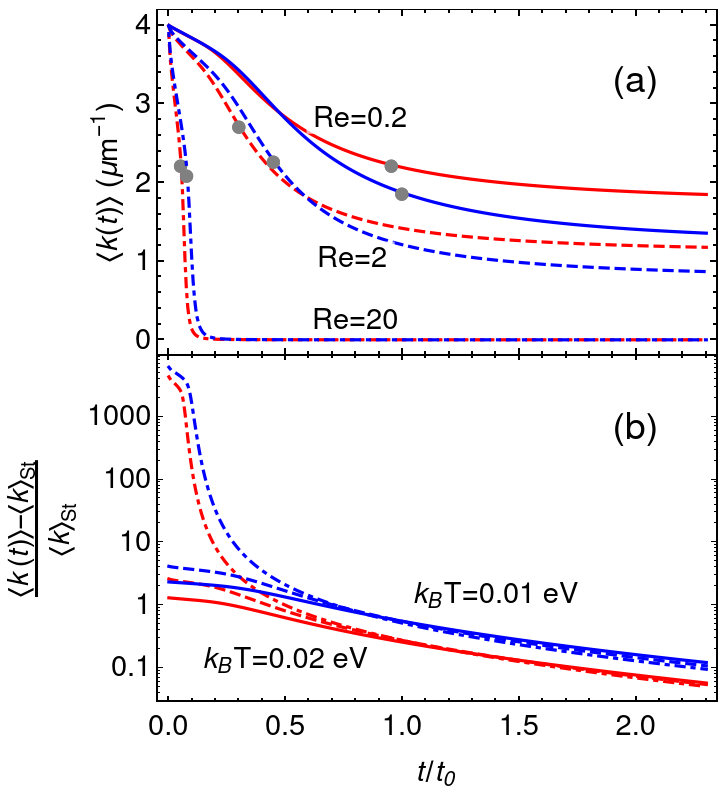}
\caption{
a - The evolution of $\langle k(t) \rangle$ for different Reynolds numbers: $Re=0.2$ (solid lines), $Re=2$ (dashed lines), $Re=20$ (dot-dashed lines) at different temperature. 
Red lines correspond to $k_{b} T = 0.02$~eV and blue lines corresponds to $0.01$~eV.
Gray dots in (a) mark the time $k_{\rm front}(t) = 0$.
Time $t_0$ defined by Eq.~(\ref{migration time}) is different for different temperatures. b - The evolution of the relative deviation of the average momentum, $\langle k(t) \rangle$,  from the stationary average momentum, $\langle k\rangle_{\rm St}$.
}
    \label{fig: universality_k}
\end{figure}

\begin{figure}
\includegraphics[width=0.85\linewidth]{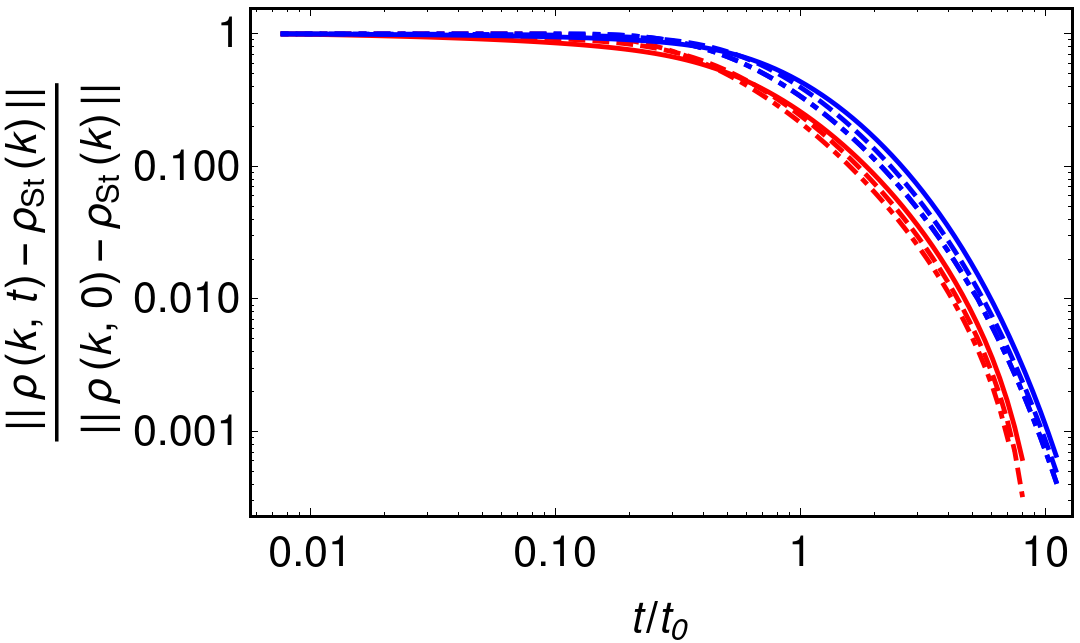}
\caption{
The evolution of the $\it{distance}$~(\ref{norm for density}) in average particle density, $\rho(k,t)$, from its stationary distribution $\rho_{\rm St}(k)$ plotted for different
Reynolds numbers: $Re=0.2$ (solid lines), $Re=2$ (dashed lines), $Re=20$ (dot-dashed lines) at different temperature. 
Red lines correspond to $k_{b} T = 0.02$~eV and blue lines corresponds to $0.01$~eV.
Time $t_0$ defined by Eq,~(\ref{migration time}) is different for different temperatures.
}
    \label{fig: universality_rho}
\end{figure}

\subsection{Dynamics of BEC formation}

We analyze the leading front $k_{\rm front}(t)$ and propose it as a measure of the evolution of the particles density $\rho(k, t)$ up to the point where $k_{\rm front}(t) = 0$. Once the ground state is reached, this measure becomes ineffective for characterizing $\rho(k, t)$, prompting us to introduce an average absolute momentum of the particles, defined as follows:
\begin{equation} \label{k average}
\langle k(t) \rangle 
=
\frac
{\int_0^{+\infty} k\rho(k,t) dk}
{\int_0^{+\infty} \rho(k,t) dk},
\end{equation}
which retains its physical meaning even when $k_{\rm front}(t)$ reaches the ground state.
In the limit of $t\rightarrow\infty$, the average momentum $\langle k(t) \rangle $ asymptotically reaches $\langle k\rangle_{\rm St}={\int_0^{+\infty} k\rho_{\rm St}(k) dk}/{\int_0^{+\infty} \rho_{\rm St}(k) dk}$, where $\rho_{\rm St}$ is defined by Eq.~(\ref{stationary solution}) (Fig.~\ref{fig: universality_k}a). Similarly to $k_{\rm front}$ the dynamics of the average momentum, $\langle k(t) \rangle$, accelerates with the Reynolds number (Fig.~\ref{fig: universality_k}a).
The evolution of $\langle k(t) \rangle$ exhibits universal evolution after the time $t_0$ defined by Eq.~(\ref{migration time}). Indeed, the evolution of relative deviation of the average momentum from the stationary average momentum, $(\langle k(t) \rangle - \langle k \rangle_{\rm St})/\langle k \rangle_{\rm St}$ does not depend on the Reynolds number for a fixed temperature (Fig.~\ref{fig: universality_k}b). 

The universal dynamics of $\langle k(t) \rangle$ suggests a corresponding universal time evolution of $\rho(k, t)$. To demonstrate this, we introduce the $\it{distance}$ between density $\rho(k, t)$ at time $t$ and stationary density $\rho_{\rm St}(k)$.
\begin{equation} \label{norm for density}
|| \rho(k,t) - \rho_{\rm St}(k) ||
=
\int\limits_0^{+\infty} | \rho(k,t) - \rho_{\rm St}(k) | dk.
\end{equation}
Figure~\ref{fig: universality_rho} shows that $|| \rho(k,t) - \rho_{\rm St}(k) ||/|| \rho(k,0) - \rho_{\rm St}(k) ||$ is independent of the Reynolds number for a fixed temperature.
This is a non-trivial result, considering the dynamics of $\rho(k, t)$ presented in Fig.\ref{fig: conservative dynamics}. Indeed, Fig.\ref{fig: conservative dynamics} clearly demonstrates a speed-up in thermalization toward the ground state as the Reynolds number increases. However, Fig.~\ref{fig: universality_rho} reveals that the overall time required for $\rho(k, t)$ to approach the stationary density $\rho_{\rm St}(k)$ — as defined by the distance (\ref{norm for density}) — is a universal measure, $t_0$, and remains independent of the Reynolds number.

\section{Conclusion}


We study the kinetics of thermalization of quantum fluids of light.
We derive the effective local form of the Maxwell–Bloch equation for the linear particle density in momentum space that takes the form of the convection-diffusion equation.
This equation describes the kinetics of the Bose gas in momentum space and reproduces the Bose-Einstein distribution as the steady-state solution in the conservative scenario.
In a particular case of a two-dimensional system with parabolic potential, the convection-diffusion equation in momentum space for quantum fluids of light resembles the well-known Bateman–Burgers equation which is widely used in fluid mechanics and describes flow propagation, shock waves, and turbulence. 
The analogy between thermalization dynamics and fluid mechanics enables us to introduce the Reynolds number for the quantum fluid of light. We demonstrate that a Reynolds number of unity corresponds to the condensation threshold in a two-dimensional Bose gas.
We obtain an approximate analytical expression describing the dynamics of the average particle density for arbitrary Reynolds numbers. This result applies to the conservative scenario, relevant to experimental platforms based on high-$Q$ optical cavities.
For Reynolds numbers greater than unity, we showed that the particles can form an analog of the shock wave in momentum space, resulting in an abrupt change in the average density at the leading front. 
Additionally, we highlight the universality in the formation of the Bose–Einstein condensate: while the dynamics of the average photon density approaching its stationary distribution depends on temperature, they show little to no dependence on the Reynolds number.

\begin{acknowledgments}
We thank M. Pr\"ufer for fruitful discussions.
A.E.S and Sh.V.Yu. thank Russian Science Foundation (project No. 20-72-10057).
P.I.V. thanks the Foundation for the Advancement of Theoretical Physics and Mathematics ``Basis''. 
A.V.Z. acknowledges support from the European Union’s Horizon 2020 research and innovation programme under the Marie Sklodowska-Curie grant agreement No 101030987 (LOREN). 
\end{acknowledgments}

\clearpage
\appendix
\begin{widetext}

\section{Derivation of the convection-diffusion equation in momentum space for quantum fluids of light} \label{appendix: derivation}
We transform the integral equation~(\ref{integral eq}) into a differential type equation (\ref{eq for linear density}).
The thermalization process is most efficient for close wave-vectors $k$ and $k'$ due to the factor $n_{\rm th}(\Delta \omega)$ in (\ref{assumption 1}) and (\ref{assumption 2}).
Moreover, the difference between the momentum $k$ and $k'$ is limited by the maximal frequency $\omega_M$, which we illustrate in Fig.~\ref{fig: q1q2_illustration}.
Thus, we introduce $q_1$ and $q_2$ as a solution of the equations $\omega_{k} - \omega_{k-q_1} = \omega_M$ and $\omega_{k+q_2} - \omega_{k} = \omega_M$ corresponding to the maximal absolute momentum that can be transferred in the elementary thermalization process 
\begin{equation} \label{q_1}
q_1 
\approx 
\frac{\omega_M}{v_{\rm gr}(k)}
+
\frac{1}{2} \frac{\omega_M^2}{v_{\rm gr}^3(k)} \frac{\partial v_{\rm gr}(k)}{\partial k},
\end{equation}
\begin{equation} \label{q_2}
q_2 
\approx 
\frac{\omega_M}{v_{\rm gr}(k)}
-
\frac{1}{2} \frac{\omega_M^2}{v_{\rm gr}^3(k)} \frac{\partial v_{\rm gr}(k)}{\partial k},
\end{equation}
where we denote the group velocity
\begin{equation} \label{group velocity}
v_{\rm gr}(k)
= 
\frac{\partial \omega_k}{\partial k}.
\end{equation}

\begin{figure}[b!]
\includegraphics[width=0.4\linewidth]{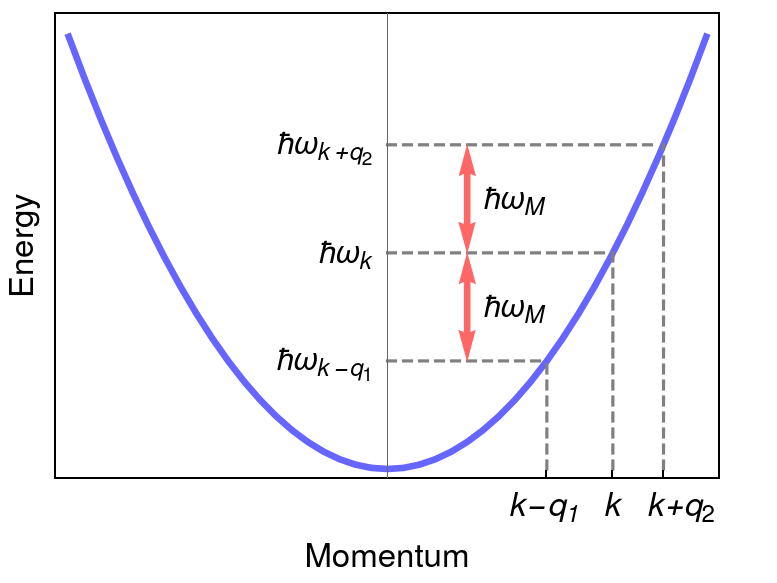}
\caption{
Schematic representation of the maximum absolute transferred momentum $q_1$ and $q_2$.
The equations determining $q_1$ and $q_2$ are $\omega_{k} - \omega_{k-q_1} = \omega_M$ and $\omega_{k+q_2} - \omega_{k} = \omega_M$.
For the illustrative purposes we show a parabolic dispersion.
}
    \label{fig: q1q2_illustration}
\end{figure}

Using Eq.~(\ref{q_1})--(\ref{q_2}) we transform the thermalization part of Eq.~(\ref{integral eq}) preserving the terms proportional to $\omega_M^2$
\begin{equation} \label{non-linear transform}
\int\limits_0^{+\infty}
(\gamma^{kk'}_{\rm therm}-\gamma^{k'k}_{\rm therm}) f(k,t)f(k',t)dk' 
=
\Gamma f(k,t)
\left[
\int\limits_0^{q_2} f(k+q,t)dq - \int\limits_0^{q_1} f(k-q,t)dq
\right]
\approx
\frac{\Gamma\omega_M^2}{2}\frac{\partial }{\partial k} 
\left( \frac{f^2(k,t)}{v_{\rm gr}^2(k)} \right)
\end{equation}
and similarly
\begin{equation} \label{linear transform}
\int\limits_0^{+\infty} 
\left[\gamma^{kk'}_{\rm therm}g(k)f(k',t)-\gamma^{k'k}_{\rm therm}g(k')f(k,t)\right]
dk'
\approx
\frac{k_BT}{\hbar}
\frac{\Gamma\omega_M^2}{2}
\frac{\partial }{\partial k} 
\left[ 
\frac{g^2(k)}{v_{\rm gr}^3(k)} 
\frac{\partial }{\partial k} 
\left(
\frac{f(k,t)}{g(k)}
\right)
\right]
+
\frac{\Gamma\omega_M^2}{2} 
\frac{\partial}{\partial k}
\left( \frac{g(k)f(k,t)}{v_{\rm gr}^2(k)} \right)
\end{equation}

The substitution of Eq.~(\ref{non-linear transform})--(\ref{linear transform}) into Eq.~(\ref{integral eq}) allow us to obtain the convection–diffusion equation in momentum space for the linear density~(\ref{eq for linear density}).

\section{Stationary solution of conservative Bateman--Burgers equation with a drag term} \label{appendix: stationary}
In this Section, we obtain the stationary solution~(\ref{stationary solution}).
The stationary conservative Bateman--Burgers equation with a drag follows from Eq.~(\ref{eq for average density})
\begin{equation}
v \frac{\partial \rho_{\rm St}(k)}{\partial (k^2)} 
+D\frac{\partial^2 \rho_{\rm St}(k)}{\partial (k^2)^2}
+b \rho_{\rm St}(k)\frac{\partial \rho_{\rm St}(k)}{\partial (k^2)} 
=
0
\end{equation}
We integrate this equation over $k^2$ and obtain
\begin{equation}
v \rho_{\rm St}(k) 
+D\frac{\partial \rho_{\rm St}(k)}{\partial (k^2)}
+\frac{b}{2} \rho_{\rm St}^2(k) 
=
C_1
\end{equation}
The boundary condition~(\ref{boundary condition}) implies $C_1=0$.
Thus, the stationary average density, $\rho_s(k)$, can be obtained from the first-order differential equation 
\begin{equation}
v \rho_{\rm St}(k) 
+D\frac{\partial \rho_{\rm St}(k)}{\partial (k^2)}
+\frac{b}{2} \rho_{\rm St}^2(k) 
=
0.
\end{equation}
The general solution of this equation is
\begin{equation}
\rho_{\rm St}(k) 
= 
\frac{2v}{b}
\frac{1}{C_2e^{vk^2/D}-1}.
\end{equation}
where $C_2$ is some constant.
From the definitions of parameters $D$, $v$ and $b$, given by Eq.~(\ref{D})--(\ref{b}), we obtain
\begin{equation}
\frac{vk^2}{D} = \frac{\hbar \alpha_{\rm cav} k^2}{k_BT} 
\end{equation}
and
\begin{equation}
\frac{2v}{b} = \frac{S}{4\pi^2}.
\end{equation}
Thus, substitution $C_2=e^{-\mu/k_BT}$ leads us to Eq.~(\ref{stationary solution}).

\section{Exact solution of conservative Bateman--Burgers equation with a drift term} \label{appendix: exact}
In this Appendix we obtain an analytical solution for Eq.~(\ref{eq for average density}) with boundary condition~(\ref{boundary condition}).

We use the Cole--Hopf transformation 
\begin{equation}
\rho(k,t)=2\frac{D}{b}\frac{\partial \ln\varphi(k, t)}{\partial (k^2)}
\end{equation}
and obtain a diffusion equation with drift for $\varphi(k, t)$
\begin{equation}
\frac{\partial\varphi(k, t)}{\partial t} 
= 
v \frac{\partial\varphi(k, t)}{\partial (k^2)}
+
D \frac{\partial^2\varphi(k, t)}{\partial (k^2)^2}
\end{equation}
with the boundary condition
\begin{equation}
v \frac{\partial\varphi(0, t)}{\partial (k^2)}
+
D \frac{\partial^2\varphi(0, t)}{\partial (k^2)^2}
=
0.
\end{equation}

Following~\cite{risken1996fokker} we use another transformation
\begin{equation}
\varphi(k, t) = e^{-vk^2/2D}\psi(k, t)
\end{equation}
we obtain a Schrodinger type equation
\begin{equation} \label{Schrodinger equation}
\frac{1}{D}\frac{\partial \psi(k, t)}{\partial t}
= 
\frac{\partial^2 \psi(k, t)}{\partial (k^2)^2}
-
\left(\frac{v}{2D}\right)^2 \psi(k, t)
\end{equation}
with the boundary condition
\begin{equation} \label{Schrodinger boundary}
\frac{\partial^2 \psi(0, t)}{\partial (k^2)^2}
-
\left(\frac{v}{2D}\right)^2 \psi(0, t)
=
0
\end{equation}

Our goal is to find propagator $G(k,t,k_1)$ such that we could obtain the solution $\psi(k,t)$ for any initial condition $\psi(k,0)$ 
\begin{equation}
\psi(k,t) = 
\int\limits_0^{+\infty} G(k,t,k_1) \psi(k_1, 0) d(k_1^2).
\end{equation}
Such a propagator should at least fulfill four following conditions
\begin{equation} \label{first condition G}
\frac{1}{D}\frac{\partial G(k,t,k_1)}{\partial t}
= 
\frac{\partial^2 G(k,t,k_1)}{\partial (k^2)^2}
-
\left(\frac{v}{2D}\right)^2 G(k,t,k_1)
\;\;
{\rm for~}
k_1 > 0,
\end{equation}
\begin{equation} \label{second condition G}
\frac{\partial^2 G(0,t,k_1)}{\partial (k^2)^2}
-
\left(\frac{v}{2D}\right)^2 G(0,t,k_1) 
=
0
\;\;
{\rm for~}
k_1 > 0,
\end{equation}
\begin{equation} \label{third condition G}
G(k,0,k_1) = \delta(k^2-k_1^2) \;\;
{\rm for~}
k_1 > 0~{\rm and}~k > 0.
\end{equation}
\begin{equation} \label{fourth condition G}
G(k,t,k_1)~{\rm is~limited~for}~{k\to+\infty}
\end{equation}
Propagator $G(k,t,k_1)$ allows us to find $\rho(k,t)$ from the initial condition $\rho(k,0)$ according to Eq.~(\ref{exact solution}).

To find the propagator $G(k,t,k_1)$ we determine the eigenfunction of the Eq.~(\ref{Schrodinger equation}) with the boundary condition Eq.~(\ref{Schrodinger boundary})
\begin{equation}
\psi_\omega(k, t) 
= 
\sqrt{\frac{v}{\pi D}} e^{-(v/2D)^2(1+\omega^2)Dt}
\sin \left(\frac{v\omega k^2}{2D}\right),
\end{equation}
\begin{equation}
\psi_{\rm st}(k, t) = \sqrt{\frac{v}{\pi D}} e^{-vk^2/2D}.
\end{equation}
One can see, that functions $\psi_\omega(k, t)$ and $\psi_{\rm st}(k, t)$ are not orthogonal.
This is because the operator in the right-hand side of Eq.~(\ref{Schrodinger equation}) with the boundary condition~(\ref{Schrodinger boundary}) is not Hermitian. 
Therefore, we cannot directly use the method described in~\cite{risken1996fokker}.
To overcome the non-orthogonality of $\psi_\omega(k, t)$ and $\psi_{\rm st}(k, t)$ we introduce a new set of functions
\begin{equation}
\bar\psi_\omega(k, t) 
= 
\sqrt{\frac{v}{\pi D}} 
\left[
e^{-(v/2D)^2(1+\omega^2)Dt}\sin \left(\frac{v\omega k^2}{2D}\right)
-
2\frac{\omega}{1+\omega^2} e^{-vk^2/2D}
\right]
,
\end{equation}
that are orthogonal to $\psi_{\rm st}(k, t)$.

We try to use the standard method~\cite{risken1996fokker} to find the propagator $G(k, t, k_1)$ 
\begin{equation} \label{third condition G}
G(k,t,k_1) 
= 
\psi_{\rm st}(k,t)\psi_{\rm st}(k_1,0)
+
\int\limits_0^{+\infty} \bar\psi_\omega(k,t)\bar\psi_\omega(k_1,0)d\omega
\end{equation}
and obtain
\begin{multline} \label{wrong G}
G(k,t,k_1) 
=
\frac{e^{-(k^2-k_1^2)^2/4Dt}e^{-v^2t/4D}}{\sqrt{4\pi D t}}
-
\frac{e^{-(k^2+k_1^2)^2/4Dt}e^{-v^2t/4D}}{\sqrt{4\pi D t}} 
+
\frac{v}{D}e^{-vk_1^2/2D}e^{-vk^2/2D}
+
\\
\frac{v}{D}e^{-vk_1^2/2D}
\left\{
{\rm sh}\left( \frac{vk^2}{2D}\right)
-
\frac{1}{2}
\left[
e^{vk^2/2D} {\rm erf}\left( \frac{vt+k^2}{\sqrt{4Dt}} \right)
-
e^{-vk^2/2D} {\rm erf}\left( \frac{vt-k^2}{\sqrt{4Dt}} \right)
\right]
\right\}
\end{multline}
This function fulfills the conditions~(\ref{first condition G})--(\ref{fourth condition G}), but gives a wrong stationary solution for $\rho$.

We note that the first two terms of Eq.~(\ref{wrong G}) fulfill the condition~(\ref{third condition G}).
Also, the first two terms and the last two terms in the right-hand side of Eq.~(\ref{wrong G}) fulfill the conditions (\ref{first condition G}) and (\ref{second condition G}) independently.
Moreover, if we change the function $(v/D)e^{-vk_1^2/2D}$ to any other function $F(k_1)$ in the last two terms of the right-hand side of Eq.~(\ref{wrong G}) the resultant propagator still fulfills the condition~(\ref{third condition G}).
To obtain the correct propagator $G(k,t,k_1)$ that preserves the total number of particles, we should replace $(v/D)e^{-vk_1^2/2D}$ in the last two terms of~(\ref{wrong G}) by $-\delta(k_1^2)$.
As a result, we obtain the propagator~(\ref{propagator}).

\end{widetext}

\bibliography{main}

\end{document}